**Semiotic problem framing: a new framework to guide students and teachers in conceptual understanding and teaching of physics**

M. Tuveri[1,3*], A. Steri[2,3], and V. Fanti[1,3]

[1] Physics Department, University of Cagliari, Cittadella Universitaria di Monserrato, 09042 Monserrato (CA), Italy

[2] Mechanical, Chemical, and Materials Engineering Department, University of Cagliari, via Marengo 2, 09123, Cagliari (CA), Italy

[3] Istituto Nazionale di Fisica Nucleare, Sezione di Cagliari, Cittadella Universitaria di Monserrato, 09042, Monserrato (CA), Italy

*Corresponding author: matteo.tuveri@ca.infn.it

**Abstract**

Problem solving in physics requires more than applying formulas: it involves describing and modeling phenomena, connecting mathematics with physics, and justifying reasoning choices. This process, known as problem framing, has been extensively studied in its cognitive and epistemic dimensions, but its semiotic aspects – how visuals, symbols, language, and metaphors shape understanding – remain underexplored. Physics relies on multiple representational modes that must be coordinated to construct meaning, and semiotics plays a central role in this integration. In this theoretical paper, we propose a new framework – the Semiotic Problem Framing – that explicitly incorporates semiotics into existing problem framing in physics. SPF highlights how students mobilize and shift across linguistic, visual, symbolic, and metaphorical resources in problem solving. For students, it offers a guide to structure reasoning and develop representational fluency; for teachers, it provides a diagnostic tool to scaffold and monitor learning processes. SPF enables analysis of reasoning patterns and error types not captured in previous frameworks, and suggests new directions for instructional design in physics education

## 1. Introduction

Effective problem-solving in physics extends beyond applying mathematical formulas; it requires understanding how mathematics connects to and reflects the physical world [1–7]. A robust epistemological framework helps students justify their decisions and link abstract reasoning to physical phenomena [8]. By *epistemological framework*, we mean the beliefs about knowledge and justification that guide how learners connect mathematical formalism to physical meaning. Applied to problem solving, this is known as problem framing. In this context, it is often conceived as a process that shapes how students interpret problems and whether they treat mathematics as a conceptual tool or as an algorithmic procedure [9–11]. Indeed, supporting students in developing thoughtful problem-framing approaches is key to fostering both conceptual and mathematical understanding [8–12].

Mathematics is central to physics, providing the language through which laws and relationships are expressed. Equations function not only as computational tools but also as representations carrying physical meaning [9]. Productive use of mathematics in physics requires blending formal operations with conceptual reasoning, enabling students to interpret equations, detect errors, and construct results that integrate physical and mathematical considerations [10]. Developing this representational fluency is therefore essential for meaningful problem solving. Instruction should not focus only on correct procedures but should help students articulate the reasoning behind their choices, connect across multiple representations, and adopt more expert-like problem-solving strategies.

However, mathematics alone does not suffice to develop expert-like competencies in physics [13-17]. Understanding concepts requires establishing new relations between knowledge elements. Indeed, physics relies on multiple modes of representing knowledge that must be interconnected to construct meaning [17-20]. These representations rely on semiotic resources, that is, different modes for representing knowledge – such as visual, symbolic, and linguistic tools (e.g., equations, diagrams, metaphors, and natural language) – that interact to create meaning. No single resource suffices; understanding arises from coordinating them across registers [21-28]. For this reason, semiotics must be considered essential to problem framing, as it contributes in shaping how students interpret tasks, construct models, and communicate reasoning.

This view aligns with a well-established line of research in physics education that conceptualizes learning as achieving fluency across a constellation of disciplinary semiotic resources. From a disciplinary discourse perspective, becoming proficient in physics entails coordinating language, mathematics, diagrams, graphs, and gestures in ways that are recognizable as legitimate within the discipline [29,30]. Within this tradition, social semiotics, e.g., the study of the development and reproduction of specialized systems of meaning making in particular sections of society [30], has been used to analyze how different representations carry specific disciplinary affordances, shaping what can be meaningfully expressed, perceived, and learned in physics [16,31]. Recent work has further emphasized purposeful transformation across modes as a central mechanism of disciplinary learning, highlighting how meaning is constructed through deliberate shifts between semiotic resources rather than through any single representation alone [32].

More research on problem framing is needed to improve how individuals are engaged both behaviorally and mentally in the learning process for actual learning to take place. Indeed, while problem framing has been extensively studied in its cognitive and epistemological dimensions [8–12], the semiotic registers have received comparatively less attention [14-16]. To bridge this gap, we propose an extended framework that incorporates semiotics into standard problem framing: the *Semiotic Problem Framing* (SPF). The rationale for this extension lies in the need to make explicit the role of semiotic resources in knowledge construction through multiple modes of representation. SPF extends rather than replaces existing frameworks, enabling a fuller account of how learners build knowledge in problem solving. In our view, this translates in considering problem framing as the integration of physics with mathematics and, notably, with semiotic resources grounded in the interaction between natural, visual, and symbolic languages. Accordingly, our study is guided by the following research question:

- How can the standard problem framing framework be extended to systematically integrate semiotic dimensions that support expert-like reasoning in physics problem solving?

The SPF framework is designed to guide both students and instructors in addressing the multifaceted nature of physics problem solving. It supports students in developing conceptual understanding and representational fluency, while providing instructors with a structured tool to analyze students' framing and reasoning. As an analytical framework, SPF enables researchers to examine how learners engage with physics problems through language, symbolism, and representation. As an instructional framework, it informs teaching strategies that scaffold students' transitions across conceptual, algorithmic, and semiotic dimensions. By explicitly linking physical reasoning with mathematical and linguistic resources, SPF fosters the development of expert-like competencies and facilitates the integration of problem framing into advanced physics education.

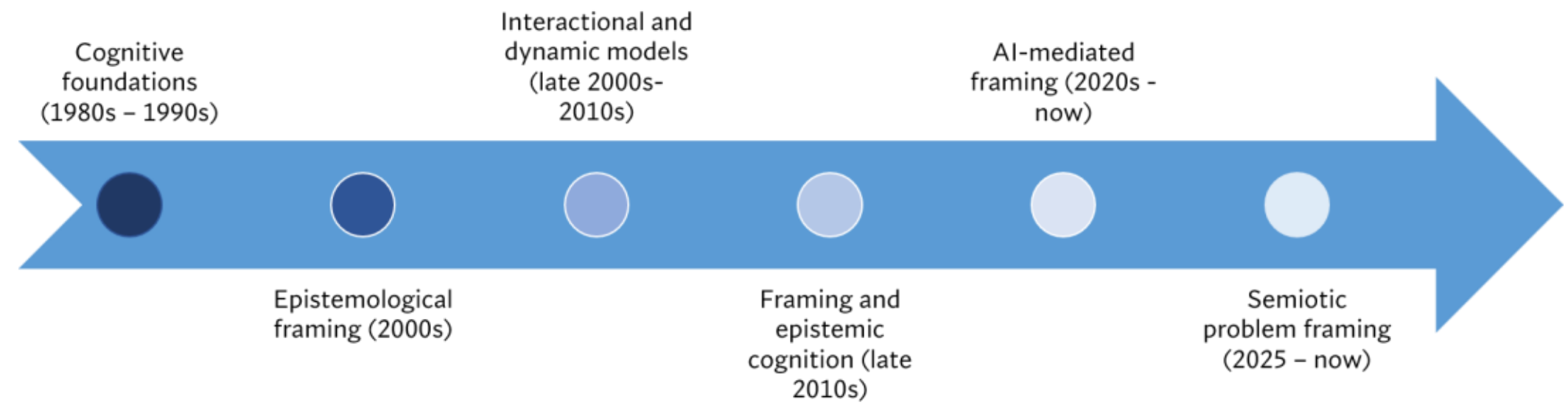

Fig. 1 The figure provides a schematic overview of key historical developments in the evolution of Problem Framing (PF) theory, from the 1980s to the present. Early work emphasized cognitive differences between novices and experts; subsequent research introduced epistemological framing and discourse-based perspectives. More recent studies integrated semiotic resources such as language, diagrams, and metaphors. In the 2020s, the rise of artificial intelligence (AI) and its implementation in education added new insights into students' problem solving, opening the route for the integration of semiotic resources into framing.

## 2. Theoretical framework

### 2.1 Background and literature review

In order to situate our contribution within the evolution of problem-framing research, Fig. 1 provides a concise historical synthesis of how the field has developed from the 1980s to the present. The figure highlights the transition from early cognitive accounts, to epistemological framing, and finally to recent semiotic and AI-mediated perspectives.

Over the past decades, problem framing in physics education has shifted from an individual cognitive process to one shaped by social and technological factors, see Fig. 1. Early cognitive studies showed that novices categorize problems by surface features, while experts focus on underlying principles – but this distinction varies with context and is not always reproducible [33]. These insights positioned problem framing as a way of organizing knowledge that influences problem-solving performance.

In the design sciences, Schön [34] introduced "problem setting" as a reflective process of redefining problems through iterative cycles. This idea was extended by Redish [35] and Hammer et al. [36] in the first decade of the 21$^{st}$ cenury in physics education through the notion of epistemological framing, where students' expectations about knowledge influence how they approach a task (see Fig. 1). Focus shifted from what students know to how they apply knowledge in context.

Bing and Redish [9,10] extended these ideas by identifying specific types of epistemological frames students use in discourse, such as physical mapping (relating a physical system to mathematics), calculation (applying mathematics to derive results), invoking authority (relying on external sources), and mathematical consistency (ensuring coherence with prior mathematical understanding). These frames reflect students' flexible and context-sensitive mental models. Fredlund et al. [16] emphasized the role of semiotic resources (e.g., equations, diagrams, gestures) in how students frame problems. Instruction that exposes expert framing practices helps students move beyond procedural thinking. As students solve problems, they switch frames dynamically, adjusting their strategies based on feedback [9,10].

When an error occurs, students' framing influences how they respond. A computational frame leads to recalculations, while a conceptual frame may prompt them to re-express the problem physically. If the error involves a flawed mapping from physics to math and students remain in a purely algorithmic frame, they may struggle to self-correct [10]. Bing and Reddish [9,10] also highlights the importance of how students justify their reasoning (or "warrants") and connect mathematical operations to physical meaning. Thus, framing serves as a methodological reference through which researchers can trace the origins of student difficulties—beyond merely tracking procedural steps.

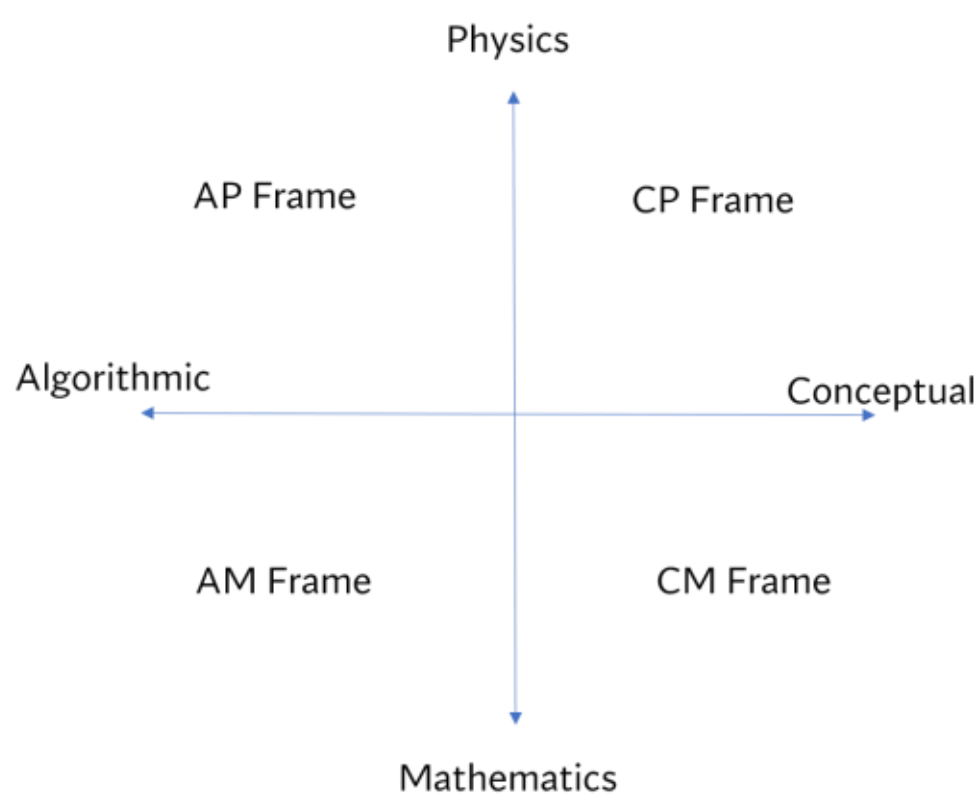

Fig. 2 Epistemological problem framing as described by Nguyen et al. [8]. The top right quadrant shows the Conceptual Physics (CP) frame, linking the Physics and Conceptual domains. In the bottom right, the Conceptual Mathematics (CM) frame connects the Conceptual and Mathematics domains. On the left side, Algorithmic Physics (AP) links the Algorithmic and Physics domains, while Algorithmic Mathematics (AM) connects the Algorithmic and Mathematics domains.

Recent studies have further refined this framework by distinguishing between specific framing types [11]. A work by Nguyen et al. (2016) [8] investigated how these frames unfold in collaborative group settings. Their analysis of undergraduate electromagnetism problem-solving revealed that epistemological frames shift continuously in real time, shaped by discourse, gestures, and tool use. The new frames in [8] help characterize the stages and transitions in students' reasoning. They are (see Fig. 2):

- CP (Conceptual Physics): engagement with physical ideas and planning based on system properties.
- AP (Algorithmic Physics): setting up equations by integrating physical knowledge with mathematical structure.
- AM (Algorithmic Mathematics): executing mathematical procedures to derive results.
- CM (Conceptual Mathematics): applying mathematical rules or shortcuts to streamline problem-solving.

These dynamics reveal that framing is a strategic, not static, stance. It involves noticing when a representation is ineffective and adapting accordingly. Recognizing the importance of epistemic cognition in framing, that is including how people understand, justify, and apply knowledge in context [33], has opened a new route to problem framing [8-10], see Fig. 1. In this context, frames can be understood as sociocultural stances embedded in disciplinary norms and institutional settings. Recognizing this helps educators to support students in adopting productive frames and to design learning environments that acknowledge the complex interplay of epistemic beliefs, resources and contexts.

Instructional strategies such as qualitative reasoning, modeling, and multiple representations can foster productive framing [34-41]. Techniques like guided decomposition and conceptual discussion support metacognitive awareness and frame-shifting. Pee et al. [42] outline three central components of framing: *seeing*, *thinking*, and *acting*. "Seeing" concerns how individuals perceive and interpret problems, often influenced by visual or conceptual representations. "Thinking" relates to internal cognitive processes shaped by prior knowledge, mental models, and metaphors. "Acting" reflects the problem-solving steps taken in response to the framed understanding. Understanding how students transition between these dimensions is essential for designing effective instruction.

More recently, the rapid emergence of artificial intelligence and its implementation in education has begun to reshape students' engagement with physics problem solving [43-47]. These tools, operating through natural language, visuals and symbols, might influence how problems are represented and solved [48-57]. While promising for scaffolding and feedback, concerns remain regarding accuracy, bias, and students' overreliance on AI outputs [58-62]. For this reason, we consider AI a contextual factor in the evolution of problem framing, but its detailed impact on learning lies beyond the scope of this theoretical work.

***2.2 Semiotics, physics learning, and problem solving***

In parallel with the development of epistemological framing models, physics education research has increasingly adopted a social semiotic perspective to investigate how meaning is negotiated through representations. Lemke's seminal work on the multiple languages of science emphasized that scientific understanding emerges from coordinated use of words, symbols, images, and actions rather than from any single representational system [63,64]. Building on this view, Airey and Linder introduced the notion of disciplinary discourse, arguing that learning physics involves mastering a critical constellation of semiotic resources that function together within specific educational contexts [29]. Semiotic resources are different modes for representing knowledge. These include natural language (both written and spoken), mathematical forms (e.g., equations), symbolic registers (e.g., vectorial and algebraic notation), and visual aids (e.g., diagrams, drawings, and pictures), as well as gestures. These diverse registers provide the foundation for constructing multiple representations of physical phenomena and systems, enabling their description, modeling, and narration [14,15,19]. Moreover, the consistent use of visual, gestural, and linguistic representations has been shown to enhance engagement in physics [18]. No single semiotic resource is sufficient on its own to fully account for physical phenomena; understanding emerges from the ability to coordinate these registers and to move flexibly between them [19]. Thus, developing representational fluency is a critical goal in physics education.

Within physics education, research in semiotics has provided a systematic framework for analyzing how representations mediate learning by foregrounding their disciplinary affordances [65]. Disciplinary affordance refers to the meaning-making potential of a representation as shaped by the norms and practices of the discipline, rather than by perceptual salience alone [31]. Studies have shown that learning challenges often stem from students' difficulty in discerning which aspects of a representation are relevant in a given context, particularly when multiple representations are involved [66-70]. From this perspective, representational competence is not merely the ability to interpret individual representations, but the capacity to coordinate and transform between them in purposeful ways [17,71]. In particular, coordinating the use of multiple semiotic registers with disciplinary affordance – what is known as multimodality – could help in enabling the construction of meaning in scientific learning and problem solving [29-31].

Subsequent studies showed how students' difficulties often arise from mismatches between representations and their disciplinary affordances, the latter characterizing how specific representations privilege certain meanings while constraining others, rather than from lack of conceptual knowledge per se [66-68]. This shapes what learners attend to and how they reason [16,31].

Concerning problem solving, recent studies have shown that being productive on it involves purposeful transformations between representations, rather than their mere use, and that such transformations are closely related to shifts in students' epistemic orientation during problem solving [32,66-68]. In this sense, social semiotics provides a natural theoretical grounding for problem framing, allowing framing dynamics to be interpreted through observable semiotic choices and transitions. Semiotic resources thus function not only as expressions of reasoning, but as active means through which students frame what kind of problem they are solving and what counts as legitimate knowledge and action within it [63,69].

Greeno's work [13] highlights how learners use mental models and situational understanding. He posited that knowledge is not static but is dynamically constructed through context-sensitive interactions. Rather than simply recalling facts, individuals apply "generative knowledge" – actively constructed, situation-specific insights. To model this, Greeno proposed the *extended semantic model*, which consists of four domains: *concrete (*physical objects and events), *model* (abstractions and simplified representations), *abstract* (concepts, principles, and laws), and *symbolic* (language, symbols, and algebraic notation). This model effectively captures how learners reason across domains, fluidly shifting between physical experiences, abstract laws, and symbolic representations.

Sherin [72] demonstrated how symbolic forms of representation, such as writing an equation, can serve as a bridge between a mathematical structure and an intuitive conceptual idea. In this way, equations function simultaneously as both formal and conceptual representations. This integration supports flexible reasoning, a deeper conceptual understanding, and the ability to translate mathematical results into physically meaningful interpretations – all hallmarks of expertise in physics [73,74].

Beyond its role in transmitting and validating knowledge, natural language constitutes the very structure of physics, which has developed its own specialized, rigorous, and dynamic discourse [74-78]. Mastery of this disciplinary language is essential for learners, yet it often presents a significant obstacle, contributing to difficulties and even alienation in physics education [79]. These challenges stem from both syntax – the formal presentation of concepts – and semantics, which is intimately tied to the structure of knowledge and model-based reasoning [21,80]. Furthermore, physics language differs markedly from everyday language, adopting impersonal and discipline-specific forms of expression that vary with context [22].

Although equations and graphs are widely recognized as central to scientific reasoning, the role of language as a representational tool has often been undervalued. As Wulff [20] points out, physics instruction is deeply language-bound, and the construction of meaning requires a blending of everyday and scientific language. Similarly, Tiberghien [14] emphasizes that language mediates meaning and facilitates transitions between different registers. Student difficulties frequently arise from ambiguities or misinterpretations of language [23-26], yet this dimension remains relatively underexplored in physics education research [22,27,28]. For this reason, natural language—whether spoken or written—plays an essential role even when a student is shifting between registers, such as from an algebraic to a vectorial form [14].

Scientific understanding is also profoundly shaped by metaphorical thought [20,28,81-87]. Conceptual metaphors project sensorimotor structures onto abstract domains, mapping embodied experiences onto otherwise inaccessible concepts [15,24]. This process creates "image schemas" that transfer knowledge across contexts, making complex ideas intelligible [88]. Metaphors function not only as simple mappings but as dynamic sign processes, co-motivated by embodied experience, cultural context, and situated use [89]. Their iconicity and occasional incongruity explain both their universality and their contextual flexibility in interpretation. Physics offers many examples: describing states as "locations" in thermodynamics, "wells" in quantum mechanics, or energy as "stuff" that can be stored, transferred, and conserved [15,20]. While metaphors are indispensable for meaning-making, students require explicit guidance to use them productively and to recognize their limitations. If taken too literally, they can lead to oversimplification and misconceptions [87]. Nevertheless, metaphors and analogies are fundamental mechanisms for constructing meaning, shaping how learners frame, narrate, and model physical phenomena [90].

The descriptive power of natural language in physics is inherently limited [20]. While it can convey qualitative relationships between variables, representing more complex functional dependencies typically requires mathematics, graphs, tables, or diagrams [91]. Scientific reasoning, therefore, often relies on mental models and thought experiments [92], with language operating alongside other semiotic resources to imbue these models with meaning. Importantly, such internal models are not encoded solely in natural language; instead, they are built from diverse personal resources, including everyday experience, prior conceptions, educational materials, and mathematical and physical knowledge acquired in both formal and informal settings, in addition to semiotic resources [93,94]. Making students aware of these semiotic and metaphorical influences can thus significantly enhance their reasoning, understanding, and learning.

## 3. Semiotic Problem Framing

### *3.1 The framework*

Taken together, the trajectory of problem framing research shows a progressive shift from early cognitive accounts, through epistemological framing, to the integration of semiotic resources, see Fig. 1. Most frameworks have primarily emphasized the interplay between mathematics and physics in scaffolding student reasoning, particularly within algorithmic, step-by-step processes. These approaches, aligned with the "acting" and "thinking" framing actions in Pee et al.'s classification [42], are valuable for tracing the dynamics of procedural problem solving. However, they do not fully capture the broader spectrum of learning processes. In particular, they either fail to make explicit or at times neglect the role of symbolic resources, already addressed in Greeno's semantic model of understanding and identified as the "seeing" domain in Pee et al.'s framework. This domain includes visual, pictorial, and metaphorical representations, which are essential for developing abstract and conceptual understandings of physics problems [14-19,40,88-95]. Despite their recognized importance, these multi-representational and epistemological domains remain underexplored in current problem framing research.

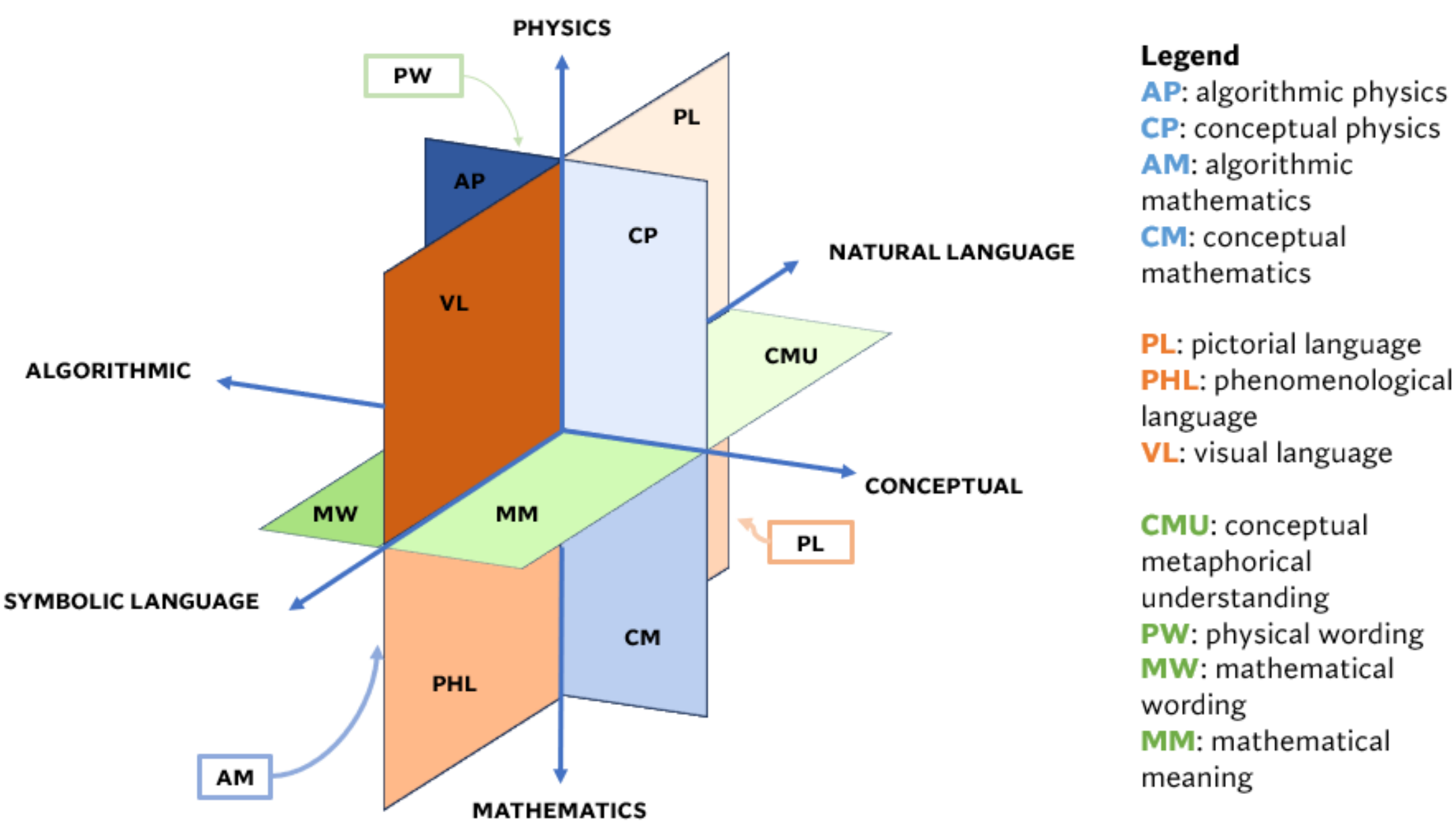

Fig. 3 The Semiotic Problem Framing (SPF) framework. The xy-plane represents the classical problem framing frameworks [8–11], defined by the physics–mathematics axis and the conceptual–algorithmic axis, generating the four epistemological frames: Conceptual Physics (CP), Algorithmic Physics (AP), Algorithmic Mathematics (AM), and Conceptual Mathematics (CM). SPF extends this framework by adding a semiotic z-axis, introducing natural and symbolic language as explicit framing resources. The intersections create new semiotic frames – Pictorial Language (PL), Visual Language (VL), Physical Wording (PW), Conceptual Metaphorical Understanding (CMU), Mathematical Wording (MW), Mathematical Meaning (MM), and Phenomenological Language (PHL). This three-dimensional structure could help in making visible reasoning processes and error types (e.g., metaphor misuse, language conflicts, representational misalignments) that could remain invisible in existing problem framing frameworks.

We argue that integrating the three cognitive dimensions – "seeing thinking, and acting" – could substantially enrich the ways in which both students and teachers frame their understanding of physics. To this end, we propose a three-dimensional framework of problem framing, the Semiotic Problem Framing (SPF). Its schematic representation is shown in Fig. 3.

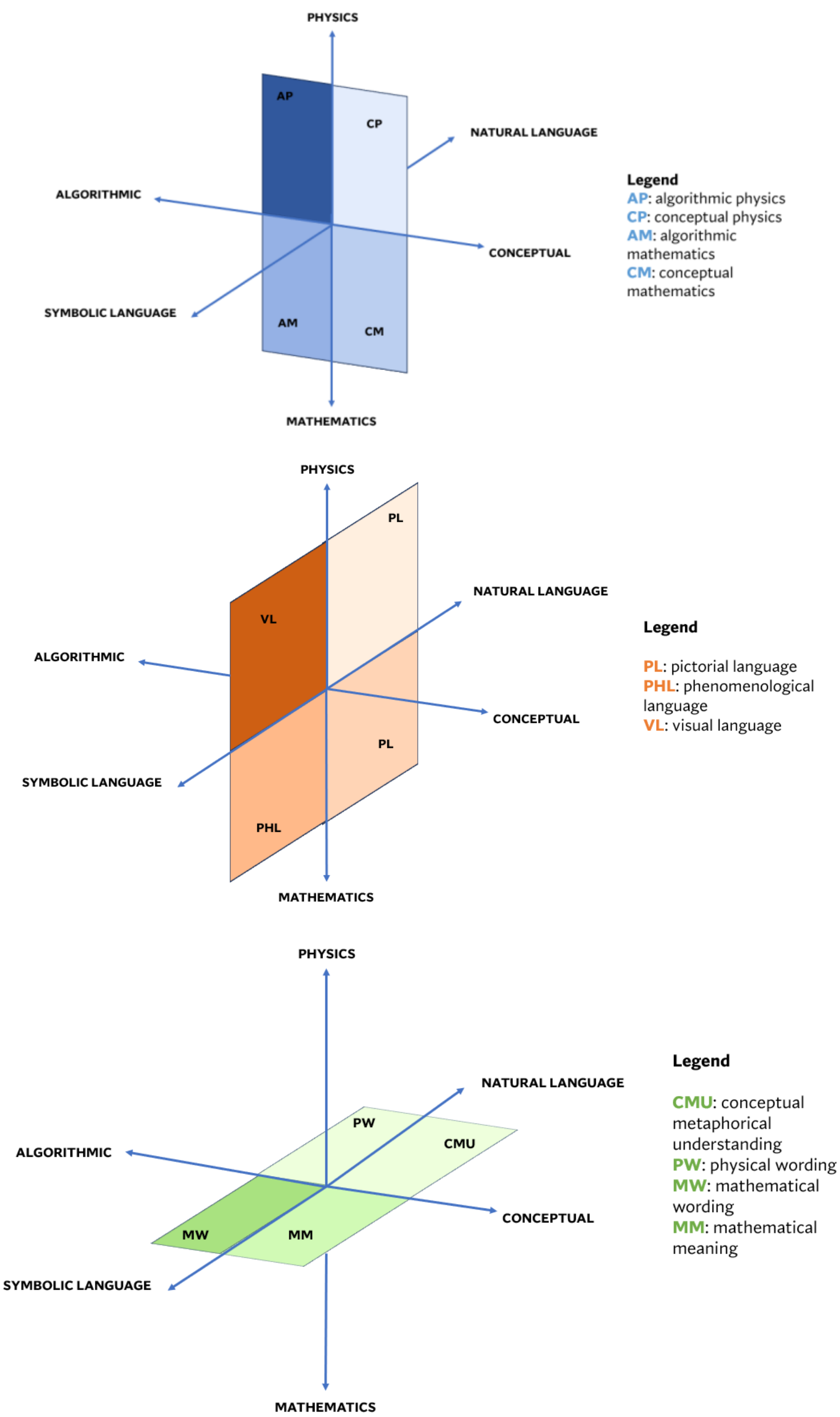

Fig. 4 Three schematic views of the Semiotic Problem Framing (SPF) framework. **(a)** The classical two-dimensional framing space from Nguyen et al. [8] and Bing & Redish [9,10], defined by the physics-mathematics and conceptual-algorithmic axes, here embedded within the SPF three-dimensional framework. **(b)** A representative plane emerging from the SPF extension with the addition of a semiotic z-axis, connecting natural and symbolic languages. **(c)** A second extension of the z-axis, generating four new semiotic frames (PW, CMU, MW, MM) at the intersections of the algorithmic-conceptual and symbolic-natural language domains. Together, these schematics illustrate how SPF integrates semiotic resources into traditional problem framing, making multimodal reasoning pathways and semiotic misalignments explicitly analyzable.

It builds on the two-dimensional xy-plane of Nguyen et al.'s and Bing & Redish's frameworks [8–10] where the x-axis represents mathematics/physics and the y-axis distinguishes conceptual from algorithmic frames (see Fig. 4a). The novelty of the SPF framework lies in the addition of a z-axis, which introduces linguistic and representational semiotic resources, encompassing natural language as well as symbolic and visual registers. We designate these as "natural" and "symbolic" languages, respectively (Fig. 3). The intersection of this new *z*-axis with the standard two-dimensional framework [8–11] generates eight new planes (or frames), which allow for a more nuanced exploration of students' framing, see Figs. 4b and 4c. We list the frames here for clarity, and elaborate on each in the following section:

- Pictorial Language (PL) and Visual Language (VL) link physics to symbolic and natural language; PL also connects mathematics with natural language.
- The Phenomenological Language frame (PHL) emerges from the intersection of mathematics and symbolic language,
- Physical Wording (PW) links physics with natural language.
- Conceptual Metaphorical Understanding (CMU) bridges the conceptual and natural language frames.
- Mathematical wording (MW) and Mathematical meaning (MM) connect the symbolic language frame with the algorithmic and conceptual frames, respectively.

Importantly, this extension does not introduce semiotics as an external add-on to problem framing, but rather systematizes insights already established within social semiotic research in physics education. By embedding disciplinary affordance, multimodality, and representational transformation directly into the framing space, SPF provides a unifying structure that connects epistemological framing models with semiotic analyses of physics learning [30,32,66-68].

Natural language connects physical and mathematical understanding through linguistic representations in both spoken and written forms. It encompasses all reasoning pathways shaped and mediated by linguistic cognitive processes. From the intersections of the algorithmic and conceptual frames with the natural language frame, two additional frames emerge, corresponding to the role of conceptual metaphors and their linguistic translation into the physical world.

Symbolic language highlights the representational role of mathematics by showing how mathematical structures connect with intuitive conceptual ideas [72]. Symbol templates capture the general form of an equation, while conceptual schemas draw on everyday reasoning patterns, such as the notion that "a whole consists of many parts." Through this blending, equations function both as tools for formal calculation and as resources for intuitive reasoning, enabling students to think in ways that are simultaneously mathematical and conceptual [10,74,75]. This integrated use of symbolic forms supports flexible problem solving and fosters a more holistic comprehension of physical systems [76].

The novelty of SPF lies not only in synthesizing prior work on problem framing and semiotics, but also in making explicit dimensions of reasoning that were previously invisible to existing frameworks. First, SPF highlights errors that stem from misapplied frames – such as overextended metaphors, misread graphs, or everyday–scientific language conflicts – that traditional frameworks cannot capture in detail. These errors, if addressed at all, tend to be subsumed within broad categories that obscure the cognitive nuances of students' reasoning. Second, by introducing a dedicated dimension for semiotic resources and their interactions with standard frames, SPF enables researchers to trace multimodal reasoning pathways, revealing how students coordinate language, visual, symbolic, and metaphorical resources across problem-solving episodes. Finally, SPF translates these analytic insights into pedagogical tools, offering teachers structured means to scaffold representational fluency by diagnosing student framings and guiding productive frame shifts. These contributions distinguish SPF from earlier frameworks, which focused primarily on the interplay between physics and mathematics in algorithmic or conceptual reasoning.

To further clarify the novelty of SPF, it is important to articulate explicitly what it captures beyond existing frameworks. The epistemological framing frameworks developed by Bing & Redish [9,10] and extended by Nguyen et al. [8] provided powerful tools to analyze how students navigate between conceptual and algorithmic approaches across physics and mathematics. However, these frameworks treat semiotic resources only implicitly, often as background tools rather than as constitutive dimensions of reasoning. SPF makes these semiotic processes explicit by:

1. Introducing a dedicated semiotic axis (z-dimension) that systematically integrates natural, symbolic, pictorial, visual, and metaphorical registers into the framing space.
2. Capturing multimodal reasoning pathways, such as how students combine diagrams with equations, or shift between everyday and scientific language.
3. Making semiotic misalignments analyzable, for instance: misuse or overextension of metaphors (e.g., energy as “stuff”); conflicts between everyday and disciplinary language; misinterpretation of graphs, diagrams, or symbolic forms.
4. Providing diagnostic value for teachers, since these error types are not easily visible within purely epistemological frames.

This comparative distinction is summarized in Table 1.

Table 1. Comparative overview of previous problem framing frameworks and the Semiotic Problem Framing (SPF). While Bing & Redish [9,10] and Nguyen et al. [8] highlight epistemological and cognitive frame dynamics, they treat semiotic resources only implicitly. SPF extends these frameworks by introducing a semiotic dimension, making multimodal reasoning pathways and semiotic misalignments (e.g., metaphor misuse, language conflicts, diagram misinterpretation) explicitly analyzable, and providing teachers with new diagnostic and pedagogical tools.

| Framework | Core dimensions | What it captures well | Limitations | SPF’s added value |
|---|---|---|---|---|
| **Bing & Reddish [9,10]** | epistemological frames (authority, calculation, consistency, mapping) | student reasoning warrants; algorithmic vs. conceptual use of math | semiotic resources treated implicitly; limited focus on multimodality | explicit semiotic axis; multimodal coordination |
| **Nguyen et al. [8]** | 2d framework (conceptual/algorithmic × physics/math) | dynamic frame shifts in group work; fine-grained epistemological analysis | no systematic account of semiotic resources; errors framed mainly as procedural or conceptual | adds semiotic frames; identifies semiotic misalignments |
| **SPF** | 3d framework (physics/math × conceptual/algorithmic × semiotic) | full integration of epistemological and semiotic dimensions | not yet empirically validated; emergence of a possible cognitive load due to the high number of frames | enables analysis of metaphor misuse, language conflicts, representational errors; offers structured pedagogical tools |

***3.2 Frames***

In the following, we describe all the frames emerging from the SPF framework. The latter incorporates Nguyen’s epistemological frames (CP, AP, CM, AM) that we list for sake of completeness. The remaining new six frames are described in detail in the following.

*Algorithmic physics (AP)*

It frames the combination of knowledge of the physical system and appropriate mathematical formulations to set up the mathematical equations of solving the problem.

*Algorithmic mathematics (AM)*

It frames the reasoning steps and the performance of algorithmic math procedure to obtain a result.

*Conceptual physics (CP)*

It frames conceptual ideas in physics, including planning a solution and analyzing relevant quantities. In this frame, planning a solution, and analyzing quantities and their role are also considered.

*CM (conceptual mathematics)*

The perception of the computation task at hand as opportunity to apply mathematical rules and properties to avoid detailed and lengthy.

*Physical wording (PW)*

The frame connects the conceptual understanding of physical phenomena to the mathematical and algorithmic computation. The connection is mediated by the natural language semiotic resource and, in particular, by the use of conceptual metaphors as a cognitive tool. PW links the AP frame [8] to natural language.

*Conceptual metaphorical understanding (CMU)*

This frame relies on the often unconscious use of metaphors to conceptualize and describe physical phenomena [96]. Metaphors and, more specifically, conceptual metaphors are a cognitive tool where we understand one idea or domain (often abstract) in terms of another (more concrete or familiar) [26]. For example, in physics, we often describe electric current as water flowing through a pipe – this maps the abstract concept of current onto the more tangible concept of fluid flow, helping learners form mental models. It involves the ability to translate physical phenomena into metaphorical representations in natural language and vice versa. This frame could be especially important for solving contemporary physics problems, such as those in quantum mechanics [97]. Recent research in physics education highlights how students struggle to relate quantum phenomena to real-world intuition and often misunderstand complex ideas like superposition or wavefunctions. Conceptual metaphors – such as visualizing quantum states as particles in potential wells or energy levels as stairs – can bridge the abstract-to-concrete gap, making quantum mechanics more accessible and aiding deeper comprehension.

*Pictorial (PL) and visual (VL) languages*

These frames focus on students' pictorial and visual representations of a physical situation. Visual representations (graphs, diagrams, simulations) support abstract reasoning (e.g., interpreting graphs) and complement verbal or symbolic forms. The pictorial representation (e.g., drawings, sketches, photographs, illustrations) helps in contextual grounding (e.g., seeing what a "mass on a spring" looks like), to visualize physical systems and anchor abstract concepts in familiar imagery and in its mathematical representation. This is why the PL frame appears two times in the three-dimensional framework in Fig. 3.

*Mathematical meaning and mathematical wording (MM, MW)*

The ability to move fluidly between conceptual and formal reasoning is a hallmark of expertise in physics [12,13,75]. Such expertise allows learners to translate mathematical solutions into meaningful physical interpretations [76–79]. These skills are particularly crucial in advanced domains, where complex equations demand not only technical accuracy but also a robust conceptual grasp of the underlying principles. Within the SPF framework, this is part of the understanding made from the use of symbolic language. This symbolic understanding is framed by two categories: mathematical meaning (MM) and mathematical wording (MW). The former concerns the meaning of mathematical operations and tasks, particularly those suitable for planning an algorithmic solution (AM in [8]). The latter refers to the ability to represent a physical situation in mathematical terms, often aligning with the application of conceptual understanding to the use of mathematical tools.

*Phenomenological language (PHL)*

The phenomenological language frame links a phenomenon to its mathematical formalization (e.g., formulas), mediated by the relevant phenomenology. For instance, in analyzing a given phenomenon experimentally, why would one consider F = ma instead of F = m/a?

Table 2. Table 1 organizes the SPF frames into three overarching dimensions – semiotic, conceptual, and algorithmic – and aligns them with Pee et al.'s [42] framing actions. For clarity, the table visually separates these categories, allowing readers to distinguish more directly which frames belong to each domain. This design highlights how SPF not only integrates existing epistemological frames but also extends them with semiotic categories that are absent from prior frameworks.

| **SPF-physics understanding dimensions** | **SPF frames** | **Framing actions (Pee et al.'s framing)** |
|---|---|---|
| Semiotic | Visual understanding<br>• Pictorial Language (PL)<br>• Visual Language<br><br>Symbolic understanding<br>• Mathematical Meaning (MM)<br>• Mathematical Wording (MW)<br>• Phenomenological Language (PHL)<br><br>Linguistic understanding<br>• Conceptual Metaphorical Understanding (CMU)<br>• Physical Wording (PW) | Seeing |
| Conceptual | Conceptual understanding<br>• Conceptual physics (CP)<br>• Conceptual mathematics (CM)<br><br>Metaphorical understanding<br>• Metaphorical (CMU, PW) | Thinking |
| Algorithmic | Algorithmic understanding<br>• Algorithmic physics (AP)<br>• Algorithmic mathematics (AM) | Acting |

More generally, we classify the SPF frames into three overarching dimensions of physics understanding, as summarized in Table 2. This classification is intended to make explicit the cognitive processes involved in framing. In doing so, we draw on research from both cognitive science and physics education, as discussed in the introduction.

The pictorial, symbolic, linguistic, and phenomenological frames are grouped within a broader dimension that we call *semiotic understanding*, since they rely on semiotic resources such as symbols, visual representations, and linguistic tools [21,22,72-80]. The metaphorical understanding contributes to what we term *conceptual understanding*, given that conceptual metaphors play a fundamental role in shaping meaning-making in physics learning [20,28,81-87]. The conceptual mathematics and conceptual physics frames also belong to this dimension. Finally, the algorithmic mathematics and algorithmic physics frames are placed within the dimension of *algorithmic understanding*, consistent with the problem framing framework of Nguyen et al. [8].

*3.3 SPF frames and framing actions*

In table 2, we also show a map between the SPF frames and the three central components of framing according to Pee et al.'s [42]. The SPF framework encourages systematic movement through three central cognitive domains – *seeing, acting*, and *thinking* – as students construct and evaluate problem-solving processes. Unlike traditional frameworks focused predominantly on algorithmic techniques [1–7], SPF supports a holistic and expert-like comprehension of problem scenarios. Prior research [8–10] has shown that students' approaches to errors are shaped by how they frame the problem. If the task is framed as a computational exercise, students may simply recheck calculations. Conversely, if the problem is understood as a translation of physical phenomena into mathematical terms, they may revisit conceptual assumptions. However, when errors stem from such translation and students persist in treating the task as merely computational, they often become stuck. As also observed in [15,16,42], existing PF frames may not sufficiently explain the entire spectrum of students' reasoning. For instance, when a student reaches an incorrect result, the approach they take to resolve the mistake depends on how they frame the problem, as also noted in [8-10]. Indeed, we suggest that students may revisit their problem-solving strategy starting from PL, VL, or PHL frames, which are related to the new action domain "seeing", see Fig. 3.

Revising calculations and conceptual aspects of both physics and mathematics ("thinking" domain) to find the error arises as a consequence to this procedure. In re-elaborating the new strategy, the thinking domain is also activated, which naturally leads to revising procedure of the semiotics register (natural language, metaphorical and visual/pictorial meanings). This differs from [8], where the revising procedure led students to come back to AM and AP frames build on mathematical and algorithmic considerations only. This contrasts with [8], where the revision process typically brought students back to the AM and AP frames, grounded primarily in mathematical and algorithmic considerations. Such a process reflects a step-by-step, algorithmic approach to problem solving that responds directly to the way the task is framed. This framing action corresponds to Pee et al.'s "acting" dimension [42] (see Table 2). In particular, the AP and AM frames involve carrying out procedures, manipulating equations, and deriving results – activities that operationalize the solution plan generated within conceptual or semiotic dimensions, or, in terms of framing actions in [42], within the seeing and thinking domains. "Acting" to implement such a plan designates the procedural enactment of a strategy rather than physical manipulation. If students frame their strategy primarily as mapping the physical situation onto mathematics, they may subsequently reconsider their understanding of the physical scenario itself.

The SPF framework expands the traditional view by offering a multidimensional tool for identifying where difficulties arise – especially in underexplored domains. In this regard, the taxonomy of error types proposed in [97] is useful and it is integrated in our framework: *displacement errors* indicate an unproductive framing of the situation, *content errors* reveal missing or inactive knowledge, and *transition errors* reflect failures to coordinate ideas across mental frames. By detailing how students shift across frames, SPF can yield new insights into how mental models interface with abstract, conceptual, and procedural aspects of problem solving.

The "seeing" domain – which links conceptual and algorithmic reasoning with metaphorical, phenomenological, symbolic, visual, and pictorial frames – is particularly relevant to this analysis. It is activated when the problem involves interpreting visual information, and it may precede or follow the "acting" phase. The sequence of these domains is context-dependent and should not be interpreted as strictly linear.

## 4. Implementation of the SPF framework in education

### *4.1 Methodological guide*

Unlike previous frameworks focused solely on procedures, SPF enables a deeper exploration of the representational challenges students face. A practical way to embed it in classroom practice is through problem-solving activities, where SPF serves as a metacognitive guide for designing strategies aligned with learning goals. Text-rich, real-world problems strengthen the link between abstract concepts and lived experiences, fostering both engagement and conceptual grounding [6]. Instructors are encouraged to analyze students' reasoning and frame transitions, enhancing their ability to provide targeted feedback and scaffold expert-like strategies. They should also examine how students' thinking evolves during these activities, with particular attention to the use of frames ("why and how a specific frame is used" and "why and how it connects to the previous and following frames") as defined by the SPF model. In collaboration with researchers, recordings of students' verbal and non-verbal work can be analyzed with SPF categories to trace how students move between conceptual, algorithmic, and semiotic frames.

Problem-solving activities should foster questioning, justification, and articulation of thought—skills essential for physics expertise [9–11]. Coordinating multiple representations (linguistic, pictorial, symbolic) deepens understanding and supports the development of expert-like reasoning. The cyclical nature of SPF also promotes reflection, allowing learners to test ideas, revise assumptions, and explore how different frames influence outcomes. Such processes strengthen both expertise and metacognition [14-16,84]. Structuring classroom activities around SPF can thus promote authentic engagement and critical thinking, with teachers playing a central role in guiding its effective use.

SPF also provides teachers with a robust lens to observe and support the full learning cycle as it unfolds in class. Students may be invited to reflect on the SPF process, identifying which frames were activated and which strategies proved most effective. The visual dimension of SPF can make abstract ideas tangible, even in traditional classrooms. Moreover, teachers can use the framework to cultivate students' self-assessment skills: the final check phase offers a structured way to evaluate coherence and consistency in reasoning, assessing their own instruction [98].

Learning to apply SPF could also serve as a methodological guide toward developing expert-like problem-solving competencies in physics. As already discussed, expert physicists do not simply execute algorithms but flexibly coordinate multiple semiotic registers – mathematical, linguistic, and visual – to interpret and model physical phenomena. SPF could guide in acquiring this expertise by explicitly structuring problem solving as an iterative movement across semiotic, conceptual, and algorithmic dimensions.

By engaging with SPF, students could become aware of their own framing processes – recognizing when they are reasoning visually, symbolically, or linguistically – and learn to regulate transitions among these registers. Such metacognitive awareness supports reflective thinking and adaptive strategy use, both essential features of expert reasoning. Consequently, SPF acts as both an analytical framework for researchers and teachers and a didactical instrument for learners, guiding them to think and act like physicists. Through iterative practice, students learn to coordinate diagrams, language, and symbols while maintaining coherence among representations. This process could help in cultivating representational fluency, epistemic flexibility, and integrated reasoning.

Finally, by making these processes visible and analyzable, SPF guides teachers to scaffold expert-like behaviors, shifting instruction from procedural correctness toward meaning-making and representational coherence. In this way, learning to use SPF becomes not merely a way to analyze problem solving, but a methodological practice that actively cultivates the very competencies it seeks to describe.

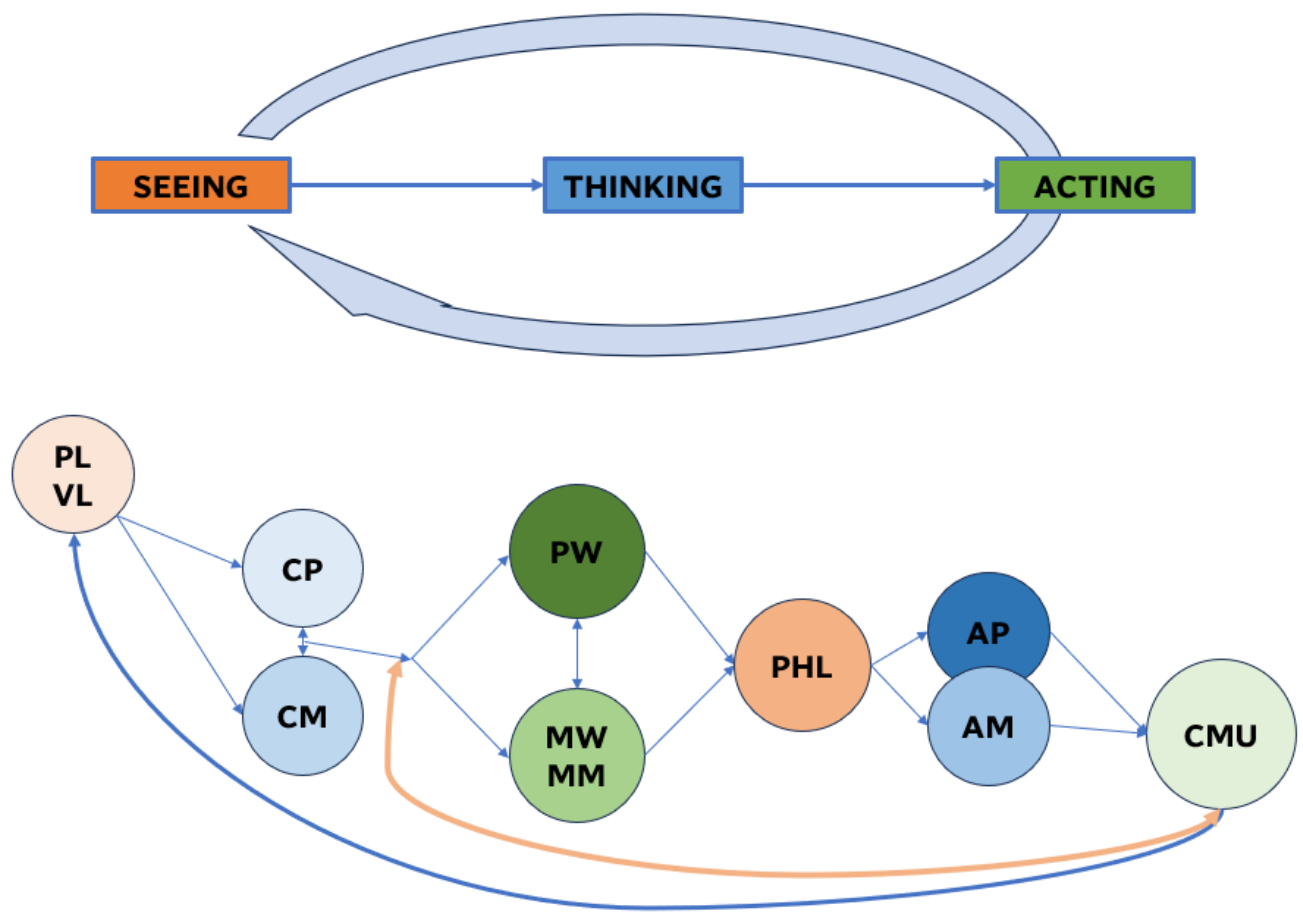

Fig. 5 A possible learning cycle aligned with the SPF framework. The upper section depicts the three cognitive domains ("seeing," "thinking," and "acting") according to Pee et al. [42]. The lower section illustrates how SPF frames map onto these domains, with arrows indicating typical frame transitions during problem solving. This representation highlights

how students often begin in the pictorial/visual frames, shift to conceptual and mathematical reasoning, and return to semiotic frames for a final consistency check. The cycle emphasizes SPF's methodological role: it provides teachers with a structured tool for observing frame shifts, diagnosing semiotic misalignments, and scaffolding expert-like reasoning in classroom practice.

### *4.2 SPF learning cycle*

This section presents a theoretical exercise designed to illustrate how the SPF framework can be applied to analyzing students' learning processes. The idea is to place the reader – whether as a teacher or as a student – inside a typical classroom scenario. As a teacher, the reader observes a student approaching a physics problem and interprets the student's reasoning through successive frame transitions according to SPF frames. As a student, the reader can instead reflect on how to use SPF as a methodological guide for problem solving and framing activities.

Authentic problem solving requires time for exploration and resolution. Rather than following a linear path from conceptual to algorithmic reasoning, students should engage in a learning cycle that highlights shifts in framing. A representation of such a cycle, aligned with SPF, is shown in Fig. 5.

From a semiotic perspective, and drawing on the framing actions described in [42], we argue that the sequence of frames – and therefore the initial approach to problem solving – should typically begin with pictorial and visual language frames. These domains provide the most immediate access to the problem, as students first rely on perceptual and representational resources before moving into more abstract conceptual and mathematical reasoning.

Why should one expect this initial approach to occur within the pictorial and visual frames? Research in physics education consistently shows that students often begin reasoning from what they can visualize – through diagrams, sketches, analogies, or concrete experiences [15,16,82,88-90,94]. These frames provide an intuitive entry point into the problem by reducing abstract complexity and activating prior knowledge. Students later shift to more abstract domains – conceptual physics, mathematical wording/meaning, and eventually algorithmic mathematics – using algorithmic, formulaic reasoning to reach solutions. While this latter approach can yield correct answers, it risks bypassing deeper conceptual understanding.

Engaging semantic frames before algorithmic execution is therefore essential: it encourages learners to interpret equations in context, test limiting cases, and assess the applicability of formal relations. In this way, equations acquire physical meaning rather than remaining isolated computational tools.

At early stages, students may also invoke metaphorical interpretations, corresponding to the activation of the conceptual metaphor frame. This promotes reframing beyond procedural steps and supports more accurate conceptualization. Subsequently, the phenomenological language frame – bridging "seeing" and "acting" through "thinking" – can help students evaluate whether algorithmic procedures align with real-world physics. This process transforms abstract representations into meaningful knowledge and fosters deeper engagement [1,5,7].

A crucial step occurs when the conceptual metaphor frame is reactivated toward the end of the process, serving as a final consistency check: verifying units, aligning mathematical form with physical meaning, and exploring whether solutions behave reasonably under boundary conditions. Represented by the orange arrow in Fig. 5, this stage emphasizes the iterative nature of learning cycles, where "seeing" is reengaged after acting and thinking. We argue that this frame also provides a valuable step prior to entering the physical wording, mathematical wording, and mathematical meaning frames.

The proposed learning cycle reflects expert-like reasoning patterns and supports the development of coherent problem-solving strategies within the SPF framework. Ultimately, the logic underlying SPF and its representation in Fig. 5 rests on the interplay of multiple semiotic frames – visual, conceptual, mathematical, algorithmic, and metaphorical – that students activate, abandon, and revisit throughout problem solving. By making these transitions explicit, educators can more effectively interpret students' reasoning, identify productive or problematic framings, and scaffold the development of expert-like competencies in physics learning.

#### *4.3. Illustrative examples*

To illustrate how the SPF framework extends traditional approaches to problem solving, we revisit a classical electromagnetism problem originally analyzed by Nguyen et al. [8]. In their study, undergraduate students worked collaboratively on the following task (see their Figure 2):

*A rectangular loop is partially inserted into a region with a time-varying magnetic field $B(t)$. Determine the induced electromotive force $\varepsilon(t)$in the loop.*

The authors reported on how students' reasoning progressed through transitions among four epistemological frames – Conceptual Physics (CP), Algorithmic Physics (AP), Algorithmic Mathematics (AM), and Conceptual Mathematics (CM) – within the two-dimensional problem-framing plane, see Fig 2. Their discourse can be summarized as follows:

1. Recalling the law of induction $\varepsilon = -\frac{d\Phi_B}{dt}$ (AP → AM frames).
2. Computing the magnetic flux $\Phi_B = \int B(t)\, dA$, while debating whether $\frac{\partial B}{\partial t}$ can be taken outside the integral (AM → CM frame).
3. Reframing the situation when one student observes that the loop's area changes with time (prompting a shift to the CP frame).
4. Completing the solution by expressing $A(x) = x \cdot y$ and obtaining $\varepsilon(t) = -\frac{d}{dt}[B(t)A(x)]$ (concluding calculations in the AM frame).

To make explicit how Nguyen et al. (2016) [8] assigned each episode to the corresponding epistemological frame, in Table 3 we provide a short explanation for each step of their analysis.

Table 3. The table shows the mapping of the four steps reported by Nguyen et al. (2016) [8] onto the corresponding epistemological frames of their two-dimensional problem-framing model. The table clarifies why each step is associated with a given frame or frame shift.

| **Step in Nguyen et al. (2016)** | **Frame(s) assigned in the original PF model** | **Explanation of the frame assignment** |
|---|---|---|
| 1. Recalling Faraday's law $\varepsilon$=−dΦ/dt | AP → AM | Students recall the physical law and translate it into a mathematical expression (AP). Immediately after, they begin performing the derivative, which is an algorithmic mathematical operation (AM). |
| 2. Computing ΦB=∫B dA and debating whether ∂B/∂t can be taken outside the integral | AM → CM | They engage in direct procedural manipulation of integrals (AM). The question about moving ∂B/∂t outside the integral reflects applying a mathematical property to simplify a computation – an instance of conceptual mathematics (CM). |
| 3. Realizing that the loop's area changes with time | CM → CP | A student recognizes a physical feature of the system (changing area), requiring reconsideration of the physical scenario and its constraints. This corresponds to conceptual physics (CP). |
| 4. Writing A(x)=x·y and deriving $\varepsilon$(t)=−d(B·A)/dt | CP → AM | Students re-express the system mathematically using physical quantities, then carry out the derivative algorithmically (AM) to obtain the final expression. |

In the authors' framework, the transitions in Table 3 describe the epistemic flow of group reasoning – from conceptual interpretation to algorithmic computation – and occasional returns to conceptual understanding.

The focus is on how students navigate between the conceptual and algorithmic dimensions of physics and mathematics.

In the SPF reinterpretation of this episode, each step is associated with a specific semiotic or epistemological frame. Such analysis reveals how language, symbols, metaphors, and representations co-construct meaning. To make this mapping explicit, we elaborate briefly on why each frame is assigned. Visual frames (PL/VL) are activated whenever students rely on sketches or diagrams; linguistic frames (PW, CMU) appear when meaning is expressed through metaphorical language; symbolic frames (MW, MM, PHL) emerge when mathematical expressions are endowed with physical or phenomenological significance. Algorithmic frames (AP, AM) correspond to the execution of formal procedures, whereas conceptual frames (CP, CM) arise when students reconsider physical assumptions or mathematical constraints. Table 4 summarizes these connections and clarifies the reasoning behind each assignment.

Table 4. Mapping of students' reasoning during the electromagnetism problem as interpreted through the Semiotic Problem Framing (SPF) framework. The table illustrates how observable actions in Nguyen et al.'s [8] episode could be linked to specific semiotic frames and highlights how students move across visual, linguistic, conceptual, and algorithmic dimensions. This reinterpretation emphasizes the role of semiotic coordination – rather than purely procedural reasoning – in shaping students' understanding of physical phenomena.

| **Stage** | **Students' observed action** | **SPF frame(s) activated** | **Semiotic interpretation** |
|---|---|---|---|
| 1 | Students draw the magnetic field and loop. | Visual Language (VL), Pictorial Language (PL) | The sketch establishes a perceptual entry point (*seeing*) and spatial relationships within the system. |
| 2 | They state that "the magnetic flux increases because the area inside the field grows." | Physical Wording (PW) | Everyday language conveys a metaphor of expansion, mapping embodied experience onto the abstract idea of flux. |
| 3 | They write $\Phi_B = \int B\, dA$ and $\varepsilon = -d\Phi_B/dt$. | Algorithmic Physics (AP) | The phenomenon is formalized through symbolic representation |
| 4 | They question whether $\partial B/\partial t$ can be moved outside the integral. | Algorithmic Mathematics (AM) and Conceptual Mathematics (CM) | Procedural manipulation dominates but paying attention to conceptual mathematical aspects of integrals; attention shifts to symbolic rules. |
| 5 | One student notes that "the area is changing with time." | Conceptual Physics (CP) + Conceptual Metaphorical Understanding (CMU) | A tension arises between linguistic and formal representations, prompting reframing. |
| 6 | They rewrite $A = xy$ and compute $\varepsilon = -\frac{d}{dt}(Bxy)$. | Mathematical Wording (MW) → Phenomenological Language (PHL) | The derivative gains physical meaning; symbolic and phenomenological registers are coordinated. |
| 7 | They check signs and physical plausibility. | Mathematical Meaning (MM) + Conceptual Physics (CP) | The final reasoning loop embodies a seeing–thinking consistency check typical of expert validation. |

The comparison between PF and SPF highlights a key difference in analytical scope. Nguyen et al.'s framework [8] effectively characterizes epistemic dynamics – how students transition between conceptual and algorithmic reasoning – but treats semiotic resources as implicit background tools. SPF, instead, makes these resources explicit analytical dimensions, allowing researchers and teachers to identify *how* representations interact to shape understanding.

In this example, the SPF analysis reveals that students' progress is guided not only by cognitive reasoning but also by the use of different semiotic resources: visual diagrams enable conceptual anchoring, metaphors connect everyday and scientific meanings, and symbolic expressions provide precision and generality.

A productive tension between different representational systems arises in students elaborates. We call it a *semiotic tension*. When students describe the phenomenon verbally – e.g., stating that "the magnetic flux increases because the area inside the field grows" – their linguistic register activates embodied metaphors rooted in everyday experience (e.g., "growth" or "expansion"). However, when they formalize the same situation as $\Phi_B = \int B\, dA$, meaning must be reconstructed within a symbolic register governed by mathematical conventions. The coexistence of these two semiotic systems, each with its own syntax and semantics, suggests the insurgence of a momentary divergence that requires reframing. Such semiotic tension serves as a trigger for deeper understanding, as students attempt to reconcile linguistic intuitions with formal expressions through explicit coordination of semiotic resources.

Similarly, what can appear as an "error" in problem solving may instead represent a case of *semiotic misframing* – that is, interpreting a situation from within an inadequate semiotic frame [9]. For instance, when students equate an "increase of flux" with an "increase of $B$" rather than with a change in the area linked to the field, the difficulty arises from remaining within a linguistic-metaphorical frame instead of shifting to a symbolic-physical one. In this view, the SPF framework allows researchers and teachers to identify the semiotic origin of reasoning difficulties, distinguishing them from purely conceptual or procedural causes. Recognizing these tensions and reframings is crucial, since it reveals how meaning in physics emerges not only from cognitive reasoning but from the dynamic interplay of multiple semiotic systems.

To further clarify how the SPF could operate, we propose another possible situation in problem solving: a student facing with the problem *"A block slides down an inclined plane with friction. Determine its acceleration."* Framed within the SPF framework, we can imagine a solution strategy like that:

1. *Pictorial and Visual Language (PL/VL):* The student begins by sketching the situation, identifying forces, and marking the incline angle, activating the *seeing* dimension.
2. *Physical Wording (PW):* They verbally describe "the block moves because gravity pulls it down the slope", translating visual to linguistic representation.
3. *Conceptual Physics (CP):* The student connects this to Newton's second law, identifying the relevant components of weight.
4. *Mathematical Wording (MW):* They express this in symbols: $F_{\text{net}} = mg\sin\,\theta - \mu mg\cos\,\theta$.
5. *Algorithmic Mathematics (AM):* They compute $a = g(\sin\,\theta - \mu\cos\,\theta)$.
6. *Phenomenological Language (PHL):* The result is checked against intuition ("if friction increases, acceleration decreases") and the student re-engages the seeing domain by comparing with limiting cases.
7. *Conceptual Metaphorical Understanding (CMU):* The student interprets friction as "resistance opposing motion," blending an embodied metaphor into physical reasoning.

This short example illustrates how SPF could capture semiotic transitions – from visual to linguistic, conceptual, and algorithmic reasoning – and how it reveals potential misalignments in reasoning (e.g., confusing the direction of friction or misusing the metaphor of "force as motion"), easily recognizable within the proposed framework.

**5. Conclusion and future perspectives**

This theoretical paper introduced a new framework for the integration of semiotics into problem framing in the context of physics problem solving. To guide this exploration, we addressed the following research questions: how can the standard problem framing framework be extended to systematically integrate semiotic dimensions that support expert-like reasoning in physics problem solving?

In response to our research question, we study and proposed the Semiotic Problem Framing (SPF) framework. Building on previous works [8–11], this framework extends the classical two-dimensional framing space – defined by the interplay between physics/mathematics and conceptual/algorithmic frames – into a three-dimensional structure. The third dimension introduces a semiotic layer largely absent from the PF literature. SPF emphasizes that learning emerges through rich, situated experiences involving dialogue, semiotic resources, and layered domains of knowledge. It aims to equip educators with tools to foster inquiry, support authentic instruction, and analyze student reasoning through meaning-centered approaches.

SPF explicitly incorporates linguistic, symbolic, and visual frames as central to meaning-making in physics. By connecting physical reasoning with symbolic operations and using metaphors as scaffolds for reframing and sense-making, the framework can help students develop more integrated and expert-like forms of understanding [14–28,33-55,74].

SPF also provides instructors with a structure to identify how students navigate across frames and semiotic registers. It supports reflective teaching, promotes inquiry-driven instruction, and enables educators to scaffold learning more effectively. Importantly, SPF also allows teachers to diagnose errors that stem from semiotic misalignments – such as overextended metaphors, misread graphs, or everyday–scientific language conflicts – that remain invisible in existing frameworks.

Nevertheless, some limitations must be acknowledged. SPF is, at this stage, a theoretical framework. Future work should focus on its empirical validation through the following possible steps: the development of a coding scheme to reliably identify semiotic frames in student discourse, gestures, and written work; inter-rater reliability tests across researchers in different educational contexts; classroom pilots at different levels, ranging from advanced secondary physics to undergraduate and graduate courses.

These empirical studies would not only test the descriptive power of SPF but also evaluate its utility as a pedagogical tool. In practice, we recommend SPF primarily for undergraduate and graduate education, where students already operate across multiple semiotic registers. A further limitation concerns the potential cognitive load associated with the high number of frames in the SPF framework. While this granularity is valuable for research purposes, it may be challenging to apply in full during classroom practice. For this reason, we suggest that simplified subsets of frames could be employed in secondary school contexts, focusing for example on pictorial, visual, and metaphorical registers, while reserving the complete framework for undergraduate and graduate levels. This staged use would reduce complexity while still fostering representational fluency and frame awareness among students.

Finally, SPF may also prove valuable for analyzing AI-mediated problem solving, where natural language, symbolic prompts, and multimodal outputs increasingly shape student reasoning [99]. While this remains beyond the scope of the present work, it represents a promising direction for future research.

To our knowledge, no comparable framework currently exists in the literature. While further empirical evidence is needed, SPF contributes a distinctive theoretical lens by systematically incorporating semiotics into PF and by offering teachers concrete strategies for observing, scaffolding, and reflecting on students' reasoning processes.

**Acknowledgments**

The authors acknowledge the anonymous Referees for suggestions and comments that help in improving the quality of the manuscript.

**Declaration of generative AI in scientific writing**

The authors declare that no generative artificial intelligence tools were used in the design, analysis, or interpretation of this study. Language editing assistance was limited to grammar and style correction using AI-assisted tools (ChatGPT-5), with all intellectual content, analysis, and interpretation solely the responsibility of the authors.

**Conflict of interest**

There are no known conflicts of interest associated with this publication, and there has been no significant financial support for this work that could have influenced its outcome.

**Orcid ID**

M Tuveri https://orcid.org/0000-0001-5686-1713

A Steri https://orcid.org/0009-0005-4963-5204
V Fanti https://orcid.org/0000-0002-4879-4183

## References

[1] Heller, P., Keith, R., & Anderson, S. (1992). Teaching problem solving through cooperative grouping. Part 1: Group versus individual problem solving. American Journal of Physics, 60, 627.

[2] Duran, M., & Dökme, İ. (2016). The effect of the inquiry-based learning approach on student's critical thinking skills. Eurasia Journal of Mathematics, Science and Technology Education, 12(12), 2887-2908.

[3] Docktor, J. L., Strand, N. E., Mestre, J. P., & Ross, B. H. (2015). Conceptual problem solving in high school physics. Physical Review Physics Education Research, 11, 020106.

[4] Leonard W. J., Dufrense, R. J., & Mestre, J. P. (1996). Using qualitative problem solving strategies to highlight the role of conceptual knowledge in solving problems. American Journal of Physics, 64, 1495.

[5] Tuveri, M., Fadda, D., Steri, A., Stefanizzi, R., Gabriele, F., Vivanet, G., Bonivento, W., Carbonaro, C. M., & Fanti, V. (2023). Promoting the learning of modern and contemporary physics in high schools in informal and non-formal contexts. Nuovo Cimento C, 46(6), 12779.

[6] Tuveri, M., Zurru, A., Fadda, D., & Saba, M. (2022). Online learning mediated by social teaching platforms: An experience from a flipped undergraduate physics course in renewable energies. European Journal of Physics, 43(5), 055703. https://doi.org/10.1088/1361-6404/ac78a6

[7] Carbonaro, C. M., Zurru, A., Fanti, V., Tuveri, M., & Usai, G. (2022). Cooperative problem solving: An experience on the training of high-school teachers and students. Giornale di Fisica, 58, 251-263. https://doi.org/10.1393/gdf/i2022-10444-x

[8] Nguyen, H. D., Chari, D. N., & Sayre, E. C. (2016). Dynamics of students' epistemological framing in group problem solving. European Journal of Physics, 37, 065706.

[9] Bing, T. J., & Redish, E. F. (2009). Analyzing problem solving using math in physics: Epistemological framing via warrants. Physical Review Special Topics-Physics Education Research, 5, 020108.

[10] Bing, T. J., & Redish, E. F. (2012). Epistemic complexity and the journeyman-expert transition. Physical Review Special Topics-Physics Education Research, 8, 010105.

[11] Thompson J. D., Modir B. & Sayre, E. C. (2016). Algorithmic, conceptual, and physical thinking: a framework for understanding student difficulties in quantum mechanics Proc. Int. Conf. of the Learning Sciences 2016.

[12] Cerny, L. T. (2012). Geometric reasoning in an active-engagement upper-division E&M classroom, PhD Dissertation Oregon State University (http://hdl.handle.net/1957/33190)

[13] Greeno, J. G. (1989). Situations, mental models, and generative knowledge. Complex Information Processing: The Impact of Herbert A. Simon, 285-318.

[14] Tiberghien, A., et al. (2005). Science and technology education at cross roads: Meeting the challenges of the 21st century. The second Conference of EDIFE and the Second IOSTE Symposium in Southern Europe.

[15] Brookes, D. T., & Etkina, E. (2012). Using conceptual metaphor and functional grammar to explore how language used in physics affects student learning. Physical Review Special Topics-Physics Education Research, 3, 010105.

[16] Fredlund, T., Airey, J., & Linder, C. (2012). Exploring the role of physics representations: An illustrative example from students sharing knowledge about refraction. Eur. J. Phys. 33 657.

[17] Linder, C., Bruun, J., Pohl, A., & Priemer, B. (2024). Relationship between semiotic representations and student performance in the context of refraction. Physical Review Physics Education Research, 20(1), 010103

[18] Duval, R. (1995). Semiosis et pensée humaine: sémiotiques registres et apprentissages intellectuels. Berna: Peter Lang.

[19] Corni, F., Fuchs, H., U., and Dumont, E. (2019). Conceptual metaphor in physics education: roots of analogy, visual metaphors, and a primary physics course for student teachers. J. Phys.: Conf. Ser. **1286** 012059

[20] Wulff, P. (2024). Physics language and language use in physics—What do we know and how AI might enhance language-related research and instruction. Eur. J. Phys. 45 02300.

[21] Bhaskar, R., & Simon, H. A. (1977). Problem solving in semantically rich domains: An example from engineering thermodynamics. Cognitive Science, 1(2), 193–215.

[22] Lemke, J. (1998). Teaching all the languages of science: words, symbols, images, and actions: La caixa conference on science education.

[23] Arons, A. B. (1997). Teaching Introductory Physics. John Wiley and Sons, Inc., New York.

[24] Lakoff G., and Johnson, M. (1980). Metaphors We Live By. The University of Chicago Press, Chicago and London.

[25] Hobson, A. (2001). The language of physics. American Journal of Physics 69, 634.

[26] Leff, H. S. (1995). Entropy and heat along reversible paths for fluids and magnets. American Journal of Physics, 63(9), 814-817.

[27] Touger, J. S. (1991). When words fail us. The Physics Teacher 29, 90.

[28] Itza-Ortiz, S. F., Rebello, N. S., Zollman, D. A. and Rodriguez-Achach, M. (2003). The Vocabulary of Physics and Its Impact on Student Learning. The Physics Teacher 41, 330.

[29] Airey, J., & Linder, C. (2009). A disciplinary discourse perspective on university science learning: Achieving fluency in a critical constellation of modes. Journal of Research in Science Teaching, 46(1), 27–49.

[30] Airey, J., & Linder, C. (2017). Social semiotics in university physics education. In Multiple representations in physics education (pp. 95-122). Springer International Publishing.

[31] Fredlund, T., Linder, C., Airey, J., & Linder, A. (2014). Unpacking physics representations: Towards an appreciation of disciplinary affordance. Physical Review Special Topics / Physical Review Physics Education Research, 10, 020129.

[32] Kapodistrias, A., & Airey, J. (2025). Multimodality and disciplinary learning: the case of purposeful transformation in the sciences. Frontiers in Communication

[33] Chi, M. T. H., Feltovich, P. J., & Glaser, R. (1981). Categorization and representation of physics problems by experts and novices. Cognitive Science, 5(2), 121–152.

[34] Schön, D. A. (1983). The reflective practitioner: How professionals think in action. New York: Basic Books. (Reprinted in 1995).

[35] Redish, E. F. (2004). in Proceedings of the International School of Physics, “Enrico Fermi” Course CLVI, edited by E. F. Redish and M. Vicentini (IOS, Amsterdam, 2004), p. 1.

[36] Elby, A., Hammer, D. (2010). Epistemological resources and framing: a cognitive framework for helping teachers interpret and respond to their students’ epistemologies. In: Bendixen LD, Feucht FC, eds. Personal Epistemology in the Classroom: Theory, Research, and Implications for Practice. Cambridge University Press, 409-434.

[37] Greene, J. A., & Ivar B. (2016). Handbook of epistemic cognition. Ed. William A. Sandoval. New York, NY: Routledge.

[38] Redish, E. F. (1994). Implications of cognitive studies for teaching physics. American Journal of Physics, 62, 796–803.

[39] Heller, J. I., & Reif, F. (1984). Describing effective human problem solving processes: Problem description in physics. Cognition and Instruction, 1, 177–216.

[40] Van Heuvelen, A. (1991). Learning to think like a physicist: A review of research based instructional strategies. American Journal of Physics, 59, 891–897.

[41] Von Korff, J., Archibeque, B., Gomez, K. A., Heckendorf, T., McKagan, S. B., Sayre, E. C., Schenk, E. W., Shepherd, C., & Sorell, L. (2016). Secondary analysis of teaching methods in introductory physics: A 50 k-student study. American Journal of Physics, 84(12), 969–974.

[42] Pee, S. H., Dorst, C. H., & van der Bijl-Brouwer, M. (2015). Understanding problem framing through research into metaphors. In V. Popovic, A. Blackler, D.-B. Luh, N. Nimkulrat, B. Kraal, & Y. Nagai (Eds.), IASDR 2015 Interplay: 2-5 November 2015, Brisbane, Australia (pp. 1656-1671).

[43] Ayeni, O. O., Al Hamad, N. M., Chisom, O. N., Osawaru, B., & Adewusi, O. E. (2024). AI in education: A review of personalized learning and educational technology. GSC Advanced Research and Reviews, 18(2), 261–271. https://doi.org/10.30574/gscarr.2024.18.2.0062

[44] Vaswani, A., Shazeer, N., Parmar, N., Uszkoreit, J., Jones, L., Gomez, A. N., Kaiser, Ł., & Polosukhin, I. (2017). Attention is all you need. Advances in Neural Information Processing Systems, 30.

[45] Yu, J. 2024. AI-driven educational reforms: Challenges and opportunities. Int. J. Educ. Technol., 47, 140–159

[46] Adiguzel, T., Kaya, M. H., & Cansu, F. K. (2023). Revolutionizing education with AI: Exploring the transformative potential of ChatGPT. Contemporary Educational Technology, 15(3), ep429. https://doi.org/10.30935/cedtech/13152

[47] Grassini, S. (2023). Shaping the future of education: Exploring the potential and consequences of AI and ChatGPT in educational settings. Educ. Sci, 13(7), 692. https://doi.org/10.3390/educsci13070692

[48] Munaye, Y. Y., Admass, W., Belayneh, Y., Molla, A., & Asmare, M. (2025). ChatGPT in Education: A Systematic Review on Opportunities, Challenges, and Future Directions. Algorithms, 18(6), 352.

[49] Montenegro-Rueda, M., Fernández-Cerero, J., Fernández-Batanero, J. M., & López-Meneses, E. (2023). Impact of the implementation of ChatGPT in education: A systematic review. Computers, 12(8), 153. https://doi.org/10.3390/computers12080153

[50] Lo, C.K. (2023). What is the impact of ChatGPT on education? A rapid review of the literature. Educ. Sci, 13(4), 410. https://doi.org/10.3390/educsci13040410

[51] Alarbi, K., Halaweh, M., Tairab, H., Alsalhi, N. R., Annamalai, N., & Aldarmaki, F. (2024). Making a revolution in physics learning in high schools with ChatGPT: A case study in UAE. Eurasia Journal of Mathematics, Science and Technology Education, 20(9), em2499. https://doi.org/10.29333/ejmste/14983

[52] Polverini, G., & Gregorcic, B. (2024). How understanding large language models can inform the use of ChatGPT in physics education. European Journal of Physics, 45, 025701.

[53] Kasneci, E., Sessler, K., Küchemann, S., Bannert, M., Dementieva, D., Fischer, F., Gasser, U., Groh, G., Günnemann, S., Hüllermeier, E., Krusche, S., Kutyniok, G., Michaeli, T., Nerdel, C., Pfeffer, J., Poquet, O., Sailer, M., Schmidt, A., Seidel, T. & Stadler, M. (2023). ChatGPT for good? On opportunities and challenges of large language models for education, Learn. Individ. Differ. 103, 102274.

[54] Tong, D., Tao, Y., Zhang, K., et al. (2024). Investigating ChatGPT-4's performance in solving physics problems and its potential implications for education. Asia Pacific Education Review, 25, 1379–1389.

[55] Urban, M., Děchtěrenko, F., Lukavský J., Hrabalová, V., Svacha, F., Brom, C., & Urban, K. (2024). ChatGPT improves creative problem-solving performance in university students: An experimental study, Computers & Education, 215, 105031

[56] Mustofa, H. A., Muhammad, R. B., & Nuraqilla W. B. G. (2024). Utilizing AI for physics problem solving: a literature review and ChatGPT experience. Lensa: Jurnal Kependidikan Fisika, 12(1), 78-97.

[57] Wang, K. D., Burkholder, E., Wieman, C., Salehi, S., & Haber, N. (2024). Examining the potential and pitfalls of ChatGPT in science and engineering problem-solving. In Frontiers in Education, 8, 1330486. Frontiers Media SA.

[58] Tossell, C. C., Tenhundfeld, N. L., Momen, A., Cooley, K., & de Visser, E. J. (2024). Student Perceptions of ChatGPT Use in a College Essay Assignment: Implications for Learning, Grading, and Trust in Artificial Intelligence. IEEE Trans. Learn. Technol. https://doi.org/10.1109/TLT.2024.3355015

[59] Tschisgale, P., Maus, H., Kieser, F., Kroehs, B., Petersen, S., & Wulff, P. (2025). Evaluating GPT-and Reasoning-based Large Language Models on Physics Olympiad Problems: Surpassing Human Performance and Implications for Educational Assessment. arXiv preprint arXiv:2505.09438.

[60] Kuhail, M. A., Alturki, N., Alramlawi, & S., Alhejori, K. (2023). Interacting with educational chatbots: A systematic review. Educ. Inf. Technol., 28, 973–1018.

[61] Krupp, L., Steinert, S., Kiefer-Emmanouilidis, M., Avila, K. E., Lukowicz, P., Kuhn, J., ... & Karolus, J. (2024). Unreflected acceptance–investigating the negative consequences of ChatGPT-assisted problem solving in physics education. In H. H. A. I. 2024: Hybrid Human AI Systems for the Social Good (pp. 199-212). IOS Press.

[62] Kosmyna, N., Hauptmann, E., Tong Yuan, Y., Situ, J., Liao, X-H., Beresnitzky, A. V., Braunstein, I., and Maes, P. (2025). Your Brain on ChatGPT: Accumulation of Cognitive Debt when Using an AI Assistant for Essay Writing Task, arXiv e-prints Art. no. arXiv:2506.08872

[63] Lemke, J. (1987). Social semiotics and science education. American Journal of Semiotics, 5(2), 217.

[64] Lemke, J. L. (1998). Teaching all the languages of science: Words, symbols, images, and actions. La Caixa Conference on Science Education.

[65] Fredlund, T. (2015). Using a Social Semiotic Perspective to Inform the Teaching and Learning of Physics. Published Doctoral thesis (Acta Universitatis Upsaliensis / Uppsala University).

[66] Fredlund, T., Airey, J., & Linder, C. (2015). Enhancing the possibilities for learning: Variation of disciplinary-relevant aspects in physics representations. European Journal of Physics, 36(5), 055001.

[67] Fredlund, T., Linder, C., & Airey, J. (2015). Towards addressing transient learning challenges in undergraduate physics: An example from electrostatics. European Journal of Physics, 36(5), 055002.

[68] Fredlund, T., Linder, C., & Airey, J. (2015). A social semiotic approach to identifying critical aspects. International Journal for Lesson and Learning Studies, 4(3), 302–316.

[69] Eriksson, M., Eriksson, U., & Linder, C. (2020). Using social semiotics and variation theory to analyse learning challenges in physics: a methodological case study. European Journal of Physics, 41(6), 065705.

[70] Eriksson, U., Linder, C., Airey, J., & Redfors, A. (2014) Introducing the anatomy of disciplinary discernment: an example from astronomy. European Journal of Science and Mathematics Education, 2(3), 167–182.

[71] Volkwyn, T. S., Airey, J., Gregorcic, B., & Linder, C. (2020). Developing representational competence: linking real-world motion to physics concepts through graphs. Learning: Research and Practice, 6(1), 88-107.

[72] Sherin, B. (2006). Common sense clarified: The role of intuitive knowledge in physics problem solving. Journal of Research in Science Teaching, 43(6), 535–555.

[73] McDermott., L. C. (1993). Guest Comment: How we teach and how students learn - A mismatch?. Am. J. Phys. 61 (4), 295–298. https://doi.org/10.1119/1.17258

[74] Distrik I.W, Supardi, Z.A.I., Jatmiko, B., Yuberti. (2021) The effects of multiple representations-based learning in improving concept understanding and problem-solving ability, J. Phys.: Conf. Ser. 1796 012044.

[75] Fulmer, G. W., Hwang, J., Ding, C., Hand, B. B., Suh, J. K. and Hansen, W. (2021). Development of a questionnaire on teachers' knowledge of language as an epistemic tool, J. Res. Sci. Teach. 58 459–90.

[76] Williamson, D. M., Xi, X., Breyer, F. J., & Educational Testing Service. (2012). A framework for evaluation and use of automated scoring. Educational Measurement: Issues and Practice, 31(1), 2–13.

[77] Martin, J. R. (1993). Literacy in Science: Learning to handle text as technology. Writing Science Pittsburgh Series in Compositions, Literacy, and Culture ed M A K Halliday and J R Martin (University of Pittsburgh Press).

[78] Koponen, I. T., Pehkonen, M. (2010). Coherent knowledge structures of physics represented as concept networks in teacher education. Science & Education, 19(3), 259-282.

[79] Tuveri, M., Gola, E., Serra, M. (2023). Teaching, Communication, and Dissemination for Society. In: Streit-Bianchi, M., Michelini, M., Bonivento, W., Tuveri, M. (eds) New Challenges and Opportunities in Physics Education. Challenges in Physics Education. Springer, Cham. https://doi.org/10.1007/978-3-031-37387-9_11

[80] Giere, R. N. (2004). How models are used to represent reality. Philosophy of science, 71(5), 742-752.

[81] Stahl, F. A. (1987). Physics as metaphor and vice versa. Leonardo, 20(1), 57-64.

[82] Corni, F. (2023). The role of metaphors in teacher education in physics. Physics Teacher Education, 3-24.

[83] Dieks, D. (2021). Analogies and metaphors in physics. Metaphors and Analogies in Sciences and Humanities, 191-213.

[84] Kokkotas, P., Vlachos I., Hatzi M., Hatzaroula V., Plakitsi K., Rizaki A., & Verdis A. (1997). The role of language in understanding the physical concepts. Nonlinear Analysis, Theory, Methods & Applications, 30(4), 2113-2120.

[85] Kramar, N. & Ilchenko O. (2021). From intriguing to misleading: The ambivalent role of metaphor in modern astrophysical and cosmological terminology. Amazonia Investiga, 10(46), 92-100.

[86] Tuveri, M., Steri A., & Fadda, D. (2024). Using storytelling to foster the teaching and learning of gravitational waves physics at high school. Physics Education, 59, 045031.

[87] Kragh, H. (2013). Big Bang: the etymology of a name, Astronomy and Geophysics, 54(2) 28-30.

[88] Lin, Q. (2024). Exploration of Conceptual Metaphor Theory in cognitive linguistics. Lecture Notes on Language and Literature, 7(1), 80-88

[89] Stampoulidis, G., Bolognesi, M., & Zlatev, J. (2019). A cognitive semiotic exploration of metaphors in Greek street art. Cognitive Semiotics, 12(1), 20192008.

[90] Fuchs, H. U. (2007). From image schemas to dynamical models in fluids, electricity, heat, and motion: An essay on physics education research. Zurich University of Applied Sciences at Winterthur.

[91] Brookes, D. T. (2006). The role of language in learning physics The State University of New Jersey

[92] Nersessian, N. J. (1999) Model-based reasoning in conceptual change. Model-Based Reasoning in Scientific Discovery ed L Magnani, N J Nersessian and P Thagard (Springer) pp 5–22

[93] Touger, J. S. (2000). The role of language in learning physics: Beyond semantics. Am. J. Phys. 68 306–7

[94] Fuchs, H. U. (2006). Modeling of Uniform Dynamical Systems. Orell Füssli Verlag, Zürich. ISBN 3-280-02771-3, Chapter 1

[95] Hestenes, D. (1987). Toward a modeling theory of instruction. American Journal of Physics, 55, 440–454.

[96] Russ, R. S., Lee, V. R. & Sherin, B. L. (2012). Framing in cognitive clinical interviews about intuitive science knowledge: Dynamic student understandings of the discourse interaction. Sci. Educ., 96, 573.

[97] Modir, B., Thompson, J. D., & Sayre, E. C. (2017). Students' epistemological framing in quantum mechanics problem solving. Physical Review Physics Education Research, 13, 020108.

[98] Brown, J. S., Collins, A., & Duguid, P. (1989). Situated cognition and the culture of learning. Educational Researcher, 18(1), 32-41.

[99] Picca, D. (2025). Not Minds, but Signs: Reframing LLMs through Semiotics. arXiv preprint arXiv:2505.17080.